# Edge localization of spin waves in antidot multilayers with perpendicular magnetic anisotropy

S. Pan[1*], S. Mondal[1*], M. Zelent[2], R. Szwierz[2], S. Pal[1], O. Hellwig[3,4], M. Krawczyk[2], and A. Barman[1]

[1]Department of Condensed Matter Physics and Material Sciences, S. N. Bose National Centre for Basic Sciences, Block JD, Sector III, Salt Lake, Kolkata 700 106, India
[2]Faculty of Physics, Adam Mickiewicz University, Umultowska 85, 61-614 Poznań, Poland
[3]Institute of Physics, Chemnitz University of Technology, Reichenhainer Straße 70, D-09107 Chemnitz, Germany
[4]Institute of Ion Beam Physics and Materials Research, Helmholtz-Zentrum Dresden-Rossendorf, 01328 Dresden, Germany
Email: krawczyk@amu.edu.pl

**Abstract**
We study the spin-wave dynamics in nanoscale antidot lattices based on Co/Pd multilayers with perpendicular magnetic anisotropy. Using time-resolved magneto-optical Kerr effect measurements we demonstrate that the variation of the antidot shape introduces significant change in the spin-wave spectra, especially in the lower frequency range. By employing micromagnetic simulations we show that additional peaks observed in the measured spectra are related to narrow shell regions around the antidots, where the magnetic anisotropy is reduced due to the $Ga^+$ ion irradiation during the focused ion beam milling process of the antidot fabrication. The results point at new possibilities for exploitation of localized spin waves in out-of-plane magnetized thin films, which are easily tunable and suitable for magnonics applications.

## I. Introduction

Spin-wave (SW) dynamics in antidot lattices (ADLs), i.e., ferromagnetic thin films with a regular two-dimensional array of holes, was extensively studied during the past decade[1,2,3,4,5]. The investigations were focused on soft ferromagnetic materials with in-plane magnetization. In this case, the properties of SWs controlled by the pattern are additionally affected by the demagnetizing field, which introduces an inhomogeneous potential for SWs.[6] In particular, at the edges of antidots the demagnetizing field decreases the internal magnetic field creating favorable conditions for SW localization.[7] The effects of ordered arrays, inhomogeneous demagnetizing field, edge modes, and configurational magnetic anisotropy in the in-plane magnetized ADL make the spectra quite complex[8,9,10] with rich physics,[11,12] but difficult for practical exploitation in magnonic devices.

In thin films with out-of-plane magnetization the forward volume SW dispersion relation is isotropic, offering better prospect for direct translation of various ideas from photonics into magnonics.[13,14,15] Moreover, a full magnonic band gap has been theoretically predicted,[16] and a fast SW guiding along a defect in perpendicularly magnetized ADL was experimentally demonstrated.[17] However, to saturate a ferromagnetic film in the out-of-plane direction, an external magnetic field exceeding its demagnetizing field is required. Ferromagnetic thin films and multilayers with perpendicular magnetic anisotropy (PMA) allow to avoid this difficulty. Ferromagnetic thin films with PMA have already found applications in magnetic recording.

However, they are rarely explored in magnonics even though they have shown some promising potential[18].

PMA originates at the interface of ferromagnetic (FM) metals (like Co, Fe) and heavy metals (HM) (like Pt, Pd) and it decreases sharply with the increase in the layer thicknesses. FM/HM multilayers can cause a cumulative increase in PMA as well as provide sufficiently large FM layer thickness for convenient detection of its magnetic response. Interestingly, the FM/HM multilayers with structural inversion asymmetry have become a recent focus of the physics community due to the emergence of interfacial Dzyaloshinskii-Moriya interaction, which helps to stabilize chiral magnetization textures and introduces SW non-reciprocity.[19]

Here, we have investigated the SW excitations in ADLs carved into Co/Pd multilayers with PMA using time-resolved magneto-optical Kerr effect (TR-MOKE) microscopy. We study antidots arranged in a square lattice with four different shapes of antidots. We have observed very rich SW spectra which depend strongly on the antidot's shape. To explain the experimental results we had to assume a region around the antidot (shell) with reduced PMA which may occur due to the focused ion beam (FIB) milling during patterning of the antidots.[20] Consequently, the demagnetizing field aligns the magnetization around the antidots in the film plane forming an in-plane domain structure. The shape of the antidot determines the domain structure and thus also determines the SW spectra. The proposed scenario is confirmed in micromagnetic simulations.

The paper is organized as follows. In the next section (Sec. II) we have defined the ADL structure, describe the fabrication and measurement techniques, and the computational approach used in the numerical simulations. In Sec. III, we describe the experimental data and support them with numerical results. The summary and possible usefulness of our findings are discussed in the last section (Sec. IV).

## II. Structure and methods

The [Co(0.75 nm)/Pd(0.9 nm)]$_8$ ML structures (shown in Fig. 1) were deposited by dc magnetron sputtering using a confocal sputter up geometry with the targets tilted and arranged in a circle around a center target (Pd) [21, 22]. The substrate, which rotates during deposition at 3 Hz, is at the focal point of the targets. The base pressure of the deposition chamber was 2 × 10$^{-8}$ mbar and the deposition was performed at 4 μbar of Ar pressure at room temperature. The ADLs are fabricated by focused ion beam (FIB) milling of the Co/Pd ML using liquid Ga$^+$ ion at 30 kV voltage and 20 pA beam current, which produces a spot size of about 10 nm. For creating the ADLs, first we create a pattern of antidots having desired shape on the ML sample. In our case, the four following shapes were investigated: triangles (T), squares (S), diamonds (Di) and circles (C), all arranged in a square lattice with slight differences between orthogonal lattice vectors, as indicated in the scanning electron microscope (SEM) images in Fig. 2(a). Subsequently, the material is milled out by exposing the patterned part to the Ga$^+$ ion beam source. Each pattern covers an area of 8 × 8 μm$^2$. The initial milling is done by using a raster scan of the focused ion beam in a single pass, which is followed by cleaning the residual materials from the ADLs in multipass (about 200 passes). The measured values of the lattice constant $a_1$ and $a_2$ along the $x$ and $y$ axis, respectively, the size of the antidots $D$ (which corresponds to the side length in T, S or Di antidots and diameter for C antidots) from the SEM images are shown in Fig. 2(a). For the four samples, the values of the structural parameters of the four ADLs are shown in Table 1.

Table 1: Structural parameters of four different ADLs as obtained from the SEM images.

| Sample | $D$(nm) | $a_1$(nm) | $a_2$(nm) | Filling Fraction |
|---|---|---|---|---|
| Circle (C) | 209 | 496 | 502 | 0.86 |
| Square (S) | 175 | 460 | 472 | 0.86 |
| Triangle (T) | 185 | 490 | 450 | 0.93 |
| Diamond (Di) | 190 | 484 | 510 | 0.85 |

All structures have similar filling fractions defined as a ratio of the ferromagnet area to the area of the unit cell $S_{Py}/a_1 a_2$.

The magnetization dynamics of the ADLs were measured by a custom-built TR-MOKE microscope in a two-colour optical pump-probe geometry [23]. The second harmonic ($\lambda$ = 400 nm, spot size ~ 1 µm, fluence = 18 mJ cm$^{-2}$) of a Ti-sapphire oscillator laser was used to excite the dynamics, whereas the time delayed fundamental laser ($\lambda$ = 800 nm, spot size ~ 800 nm, and fluence 2.5 mJ cm$^{-2}$) was used to probe the dynamics. The two beams were made collinear before being incident on the sample through a single microscope objective with numerical aperture of 0.65. A piezo-electric scanning $x$-$y$-$z$ stage is used to locate the pump and the probe beams at the central region of each array during the measurement. The back-reflected probe beam is collected by the same microscope objective and is analyzed to measure the polar Kerr rotation by an optical bridge detector as a function of the time-delay between the pump and probe beams. A bias magnetic field ($H$ = 2.23 kOe) is applied at a small angle (~ 10°) to the surface normal of the sample.

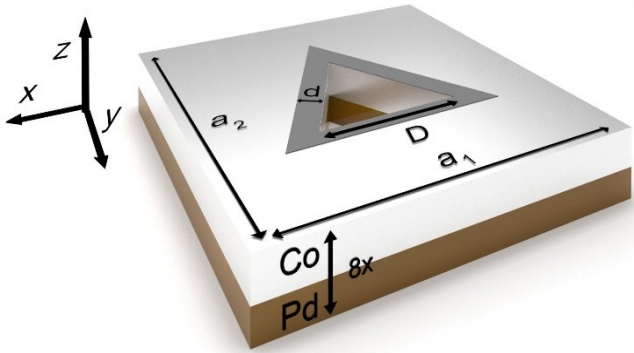

**Fig. 1.** A schematic of the multilayered ADL structure composed of Co and Pd layers under investigation and the definition of the coordinate system used throughout the paper. Here the antidot has a triangle shape with side length D and the shell with width d around it with reduced PMA is marked by the grey color. $a_1$ and $a_2$ are the $x$ and $y$ lattice constants of the antidot array, respectively.

To interpret the experimental results we performed micromagnetic simulations.[24] We assumed homogeneous material across the thickness with the effective parameters taken from the literature,[21] $K_u$ = 4.5 x 10$^5$ J/m$^3$, $M_S$ = 0.81 x 10$^6$ A/m, exchange constant $A$ = 1.3 x 10$^{-11}$ J/m, and gyromagnetic ratio $\gamma$ = 187 x 10$^9$ rad GHz/T. The damping was neglected in the simulations. The simulations have been performed for the unit cell with dimensions obtained from the SEM images after applying periodic boundary condition along the $x$- and $y$-axis. The discretization for all structures was fixed to 1 x 1 x 4.4 nm$^3$.

The stabilization of the magnetization configuration was obtained in the following steps. We assumed an uniform shell with constant width and shape corresponding to the shape of the respective antidots, where the magnetic anisotropy was reduced to zero. The initial magnetization of each cell was set to a random magnetization, except the shell, where the orientation of the magnetization was forced into either a vortex state, or alternatively into a head-to-head and tail-to-tail domain wall state in the shell around the antidot. In the following step the system was relaxed at magnetic field equal to 2.23 kOe. To excite SWs we use a microwave magnetic field $h_{mf}$ directed along the $x$ axis. $h_{mf}$ was uniform in each unit cell with the shape of the sinc function in the time domain with cutting frequency $f_c$ = 25 GHz and amplitude 15 Oe. The assumed conditions are close to the detection of SWs with TR-MOKE, where the width of the probe beam is larger than the unit cell.

**Results**

Time-resolved traces from the TR-MOKE measurements are presented in Fig. 2(b), while their fast Fourier transformed (FFT) spectra are presented in Fig. 2(c). The vertical dashed line indicates the frequency (10.6 GHz) of the SW measured from the unpatterned ML film. In ADLs, the most intense peak (marked as '1') changes only slightly with the change in the antidot's shape (10.30, 10.80, 10.13 and 10.80 GHz for C, S, T and Di, respectively), but the number and position of the peaks with lower intensity vary significantly. For the C-antidot there are two peaks relatively close to each other at 8.25 and 10.30 GHz. For the S-antidot also only two peaks (7.25 and 10.80 GHz, marked as 3 and 2, respectively) appear. However, for the T-antidot three clear peaks appear, two with smaller intensity at 2.58 and 7.73 GHz, and an intense peak at 10.13 GHz. The richest spectra is measured for the Di antidot, which consists of a set of five peaks (5.15, 6.70, 8.25, 9.28 and 10.82 GHz) with varying intensities, out of which two peaks have comparable intensity.

To elucidate the type and origin of the measured SW excitations we have performed micromagnetic simulations in samples with dimensions as extracted from the SEM images in Fig. 2(a) at an applied magnetic field of 2.23 kOe. For all four ADLs we obtained a single peak spectrum with only slight changes of the peak position between ADLs with different antidot shape (10.21, 10.24, 10.21 GHz for C, S and D, respectively). The spatial distribution of the SW amplitude extracted from the micromagnetic simulations reveals the fundamental mode, i.e., the mode with the in-phase magnetization oscillations in the entire sample. The slight variation of the fundamental mode frequency with the variation in the antidot shape is reasonable, considering that the ferromagnetic material occupies about 85% of the area, independent of the structure. However, these simulation results do not explain the experimental spectra in Fig. 2(c). This indicates the presence of a missing factor crucial for the observed multi-mode SW excitations, which has not been included in the simulation.

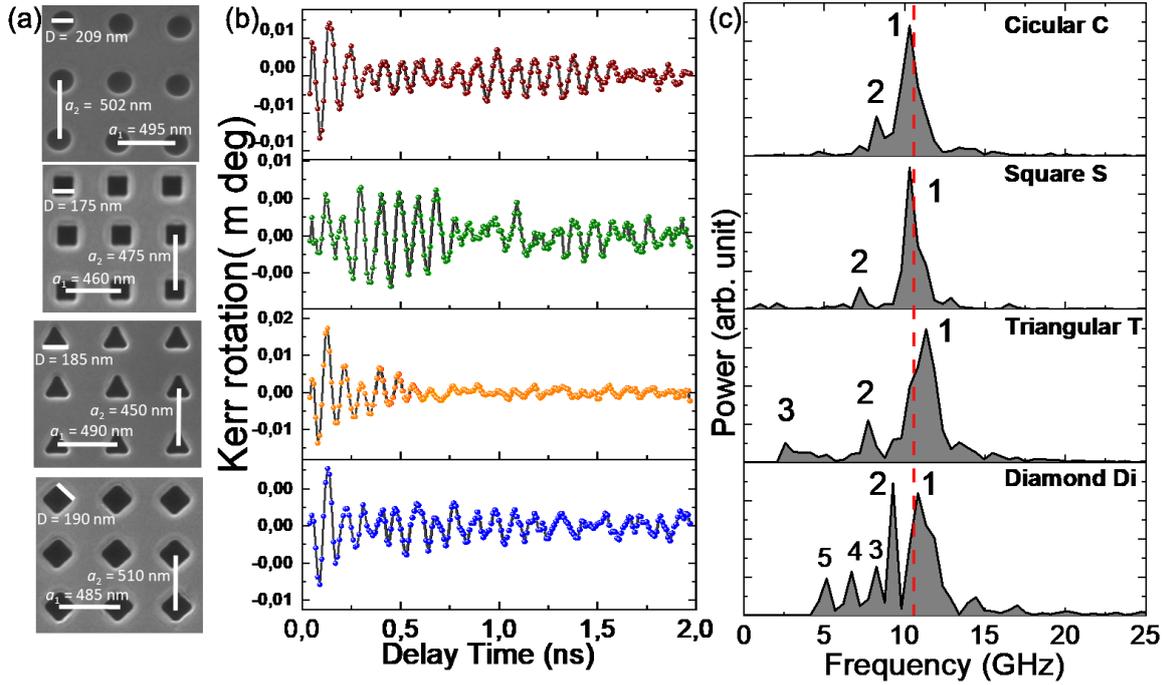

**Fig. 2.** (a) The SEM images of the four types of ADLs investigated in the paper. (b) TR-MOKE signal after subtracting a bi-exponential background and (c) its FFT spectra at $H$ = 2.23 kOe for ADLs with different shapes of the antidots: circular C, square S, triangles T and diamond Di. The vertical red-dashed line points at the SW in the plain ML film.

The presence of additional peaks below the frequency of the fundamental mode points at the lowering of magnetic anisotropy or saturation magnetization, as both these two factors can decrease the frequency of SWs. A previous study indicated[21] that these two magnetic properties near the antidot edges in PMA multilayers can be strongly affected during the fabrication process. Indeed, in the literature we have found confirmation that light ion irradiation can affect the magnetic properties in PMA multilayers. In ref. [25] the decrease of PMA due to interface roughening and alloy formation caused by ion irradiation in Co/Pt ML systems has been demonstrated. The FIB irradiation has also been used to control the magnetic anisotropy in multilayers with PMA.[26,27] In our case, during the process of ion milling of the Co/Pd MLs for the creation of the antidots some scattered ions may have penetrated through the sidewalls of the antidots, causing a degradation of the magnetic properties over a narrow rim-like region around the antidots. These rim-like region around the antidots (shells) cover a small fraction of the ADL area. For instance, a shell with width, $d$ = 10 nm around a C-type antidot will constitute only 3.2% of the ML area. The interesting question is, whether the MOKE signal from such a small area collected during TR-MOKE measurement can have significant intensity as compared to the fundamental mode occupying over 80% of the surface area? To check this, we performed simulations with $K_u = 0$ in the $d$ = 10 nm wide shell, while keeping other parameters unchanged.

The results are shown in Fig. 3. The SW spectra [Fig. 3(a)] is very close to the TR-MOKE result presented in Fig. 2(b). There are two intense peaks, one with higher intensity at 10.2 GHz (marked as ii, close to the resonance peak in the plain ML), the other one (i) at 8.3 GHz. Other peaks of smaller intensity also appear, but as they were not detected in the measurements they will not be analyzed here. The ground state magnetic configuration in the unit cell is shown in Fig. 3(b), where the white color means the out-of-plane magnetization, and the color indicates orientation of the magnetization in the film plane. It becomes clear, that lack of the PMA in the shell region allows the demagnetizing field to align the magnetization into an in-plane closure domain structure (like a vortex) around the antidots. The extracted amplitude profiles and phase of SWs shown in Fig. 3(c-f) demonstrate, that the low frequency mode (i) has its amplitude concentrated in the area of the decreased anisotropy, known as the edge SW mode. The mode (ii) at 10.2 GHz has its amplitude spreading over the whole area, but with an out-of-phase contribution from the magnetization precession in the shell. Importantly, magnetization precession in the shell generates much larger amplitude of the dynamical out-of-plane component of the magnetization than the magnetization in other part of the ML which is normal to the film plane or slightly tilted from this direction. As the TR-MOKE signal is proportional to the out-of-plane component of the dynamic magnetization (polar MOKE geometry), the peak related to the SW edge mode appears with significant intensity as compared to the intensity of the fundamental mode. Moreover, the collective response of the fundamental mode in the measurements is additionally decreased due to the negative contribution of the out-of-phase oscillations in the shells, bringing the fundamental mode intensity level close to the edge mode intensity level.

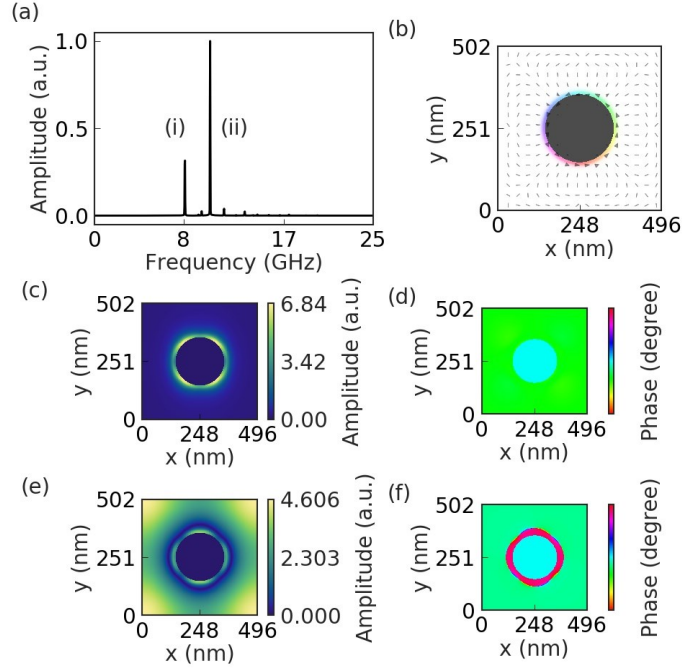

**Fig. 3.** (a) The simulated spectra of SWs in the C antidot with the zero PMA value in the shell of 10 nm width around the antidot, with $H = 2.23$ kOe. (b) The static configuration of the magnetization stabilized in the structure. (c, e) The amplitude of the out of plane component of the dynamical part of the magnetization, and (d, f) the phase of high intensity SWs at 8.3 and 10.2 GHz, indicated as i and ii in (a), respectively.

To clarify the influence of the relative intensities of the edge and fundamental SW modes, we performed simulations for different $d$ values for the sample C. We found a nonlinear dependence (Fig. 4) and observed a mode crossover of the two most intense peaks with increasing $d$. At small $d$, the higher frequency peak appears with higher intensity, while for larger $d$ intensity becomes higher in the lower frequency mode, the crossover occurs at $d \approx$ 12 nm. Interestingly, we do not observe a crossing of the frequency peaks in Fig. 4, rather an anti-crossing is observed, which indicates a hybridization of the edge and fundamental mode, justifying the mixed mode profile in Fig. 3(e). We need to stress, that the interaction between these modes depends on many parameters, including the profile of the anisotropy increase from the antidot edge and the variation of other magnetic parameters in the shell, which could not be experimentally measured. Consequently, the study of these dependences are out of the scope of this paper.

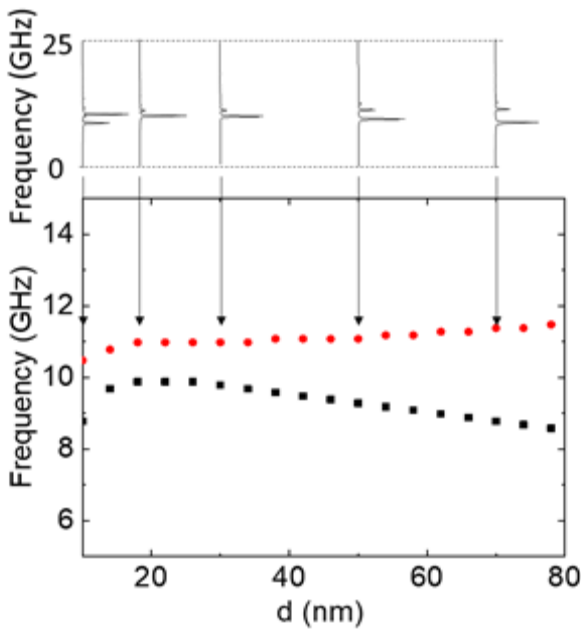

**Fig. 4.** Simulated spectra (top panel) and the frequency of the most intense modes (bottom panel) as a function of the width of the shell $d$ around the antidots in the sample C.

The numerical spectra for the S, T and Di samples with 10 nm shell having zero PMA, together with the static magnetization configurations, and selected SW profiles are shown in Fig. 5. Similar to the C antidots, the static magnetization forms closure domains in the shells, aligned tangentially to the antidot edges. Due to different shapes, their magnetization configurations are different: for the S and Di there are four domains along the antidot edges with four domain walls at the vertices of the square. For T there are 3 domains separated by domain walls.

The calculated spectrum for the S-antidot has two resonant peaks of high intensity at 8.33 GHz (i) and 10.54 GHz (iii), along with a number of peaks having smaller intensity [Fig. 5(a)]. The spectrum is very close to the experimental result shown in Fig. 2(c). The lower frequency mode [8.33 GHz in Fig. 5(a)] (i) has an amplitude in the shell with maximum at the vortices of the square antidot. The most intense peak (iii) is the fundamental mode, while the modes of low intensity at 9.64 and 9.8 GHz (ii) are the excitations with the amplitude concentrated in the domains along the antidot side walls. A qualitative agreement between the simulated and the experimental results [Fig. 2(c)] is obtained also for the T-antidot structure [Fig. 5(b)]. There are two moderately intense resonant peaks at 6.95 GHz (i) and 7.32 GHz, two very weak peaks around 8.8 GHz, one reasonably intense peak at 9.7 GHz (ii) all of which are well separated from the most intense fundamental mode at 10.21 GHz (iii). According to the calculated profiles the two SWs of lowest frequency have amplitude concentrated in the domain walls at the vertices (like the mode (i) at 6.95 GHz with the amplitude at the top vertex shown in the inset). The modes at 8.8 and 9.73 (ii) GHz are concentrated in the domains along the triangle antidot sides. We checked that an increase in $d$ from 10 to 20 nm results in the low frequency modes shifting toward the fundamental mode, which makes the spectra even closer to the experimental one. Interestingly, in Ref. [28] three-peak SW spectra were detected for the in-plane magnetized isolated triangular rings made of Py. Although the width of the rings (380 nm) was almost 40 times more than the shell assumed in our ADLs, the spectra and the mode profiles are qualitatively very similar.

The spectra for Di-ADL is shown in Fig. 5(c). The most intense peak (iii) related to the fundamental mode is at 10.41 GHz. The mode of lowest frequency (i) is at 7.37 GHz and it corresponds to the vertex excitation. The low intensity mode (ii) at 8.6 GHz is a SW in the domains of the shell. Overall the spectrum of Di-ADL is similar to the S-ADL spectra. Because, the shape of the antidots in both structures are almost the same, the differences between the S and Di spectra can be related to the effects of interaction between different antidots in the lattice. This suggests collective dynamics, not only of the fundamental mode, but also of the edge SW modes. Nevertheless, the simulated spectrum of Di-ADL structure does not match with the experimental spectrum shown in Fig. 2(c), where 5 resonant modes are clearly visible. An increase of shell width in the simulation changes the relative intensities of the two modes, essentially retaining the two peak spectra, as can be observed in Fig. 6(a) for a Di-ADL structure with $d$ = 20 nm.

To underpin the origin of the discrepancies between the experimental and the simulated spectra for the Di-ADL type, we revisit the ground state magnetization configuration and the SW spectra measured in ring structures of square, diamond and rectangular shapes made of soft ferromagnetic materials.[29,30,31] Those reports state that the closure domain is a magnetic ground state configuration, but at remanence other magnetic patterns can also form. The latter include configurations with the head-to-head and tail-to-tail domain walls trapped in the vertices of the shells around the antidots. The change in the magnetic configuration is manifested in the variation of the SW spectra. Apart from having much smaller width of the shells assumed in our ADLs as compared with the reported nanorings (tens of times narrower), different magnetic configurations realized in the shells probably lead to an increase in the number of the SW modes in our ADLs. Indeed, during measurements we do not control the magnetization alignment in the shells, closure domain or configurations with the head-to-head and tail-to-tail domain walls at the vertexes can be stabilized at remanence or at small magnetic field. In Fig. 6(b), we show the SW spectra for Di-ADL with $d$ = 20 nm shell, with one head-to-head and tail-to-tail domain walls in opposite vertexes of the diamond. Interestingly, this spectra is different from the spectra in Fig. 6(a). In Fig. 6(b), there are additional resonant modes at 6.2 GHz (i) and above 9.6 GHz (ii), with amplitude concentrated at vertices with head-to-head and tail-to-tail domain walls, and in the domains respectively, making this spectra closer to the experimental spectra.

The difference between experimental and numerical results can be related also to the tilted external magnetic field form the surface normal orientation. For homogeneous ML and ADLs, this has a minor effect. However, in the shell with the in-plane closure domain structure, the magnetic field orientation breaks the domain symmetry which can result in increasing number of modes. This scenario found confirmation in the results of micromagnetic simulations presented in Fig. 6(c), where the SW spectra for the Di-ADL structure with the magnetic field tilted by 10º from the surface normal direction towards the *x*-axis is presented. The closure domain structure was relaxed in the shell and we found a number of low intensity resonances at lower frequencies (below 9 GHz). The mode (i) of the largest intensity at 10.2 GHz, shown in the inset of Fig. 6(c), can be considered as a fundamental mode hybridized with the asymmetric edge mode, whose amplitude is concentrated in the domains with and without nodal lines along the opposite antidot edges. Also the other simplifications made in our model can result in discrepancies between simulation and experimental results. As mentioned above, in the real sample the value of anisotropy and saturation magnetization may vary continuously, while in the simulation these quantities are assumed constant. However, these properties could not be experimentally detected so far, and hence further experimental and theoretical investigations are necessary to clarify the validity and relative contributions of the proposed effects.

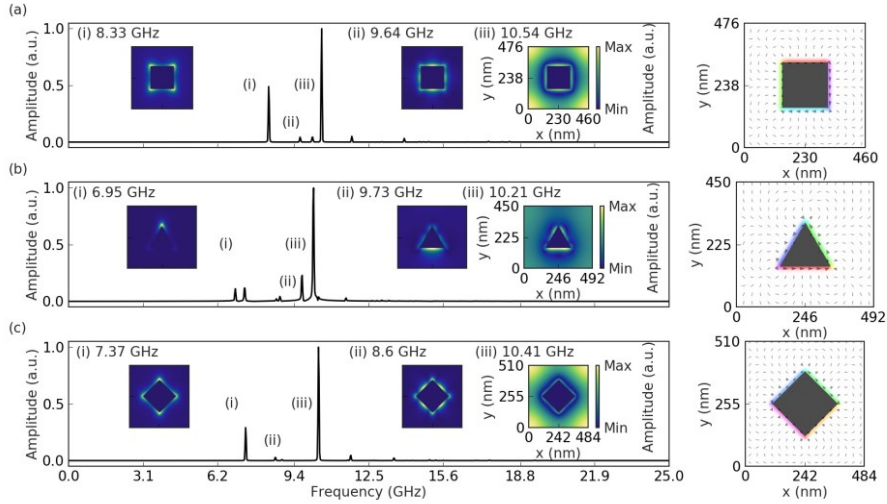

**Fig. 5.** Simulated SW spectra for the S-, T- and Di-ADL structure with $d = 10$ nm with the respective static magnetization configurations and the profiles of the most intense lines in the spectra. The out of plane magnetic field $H = 2.23$ kOe was used.

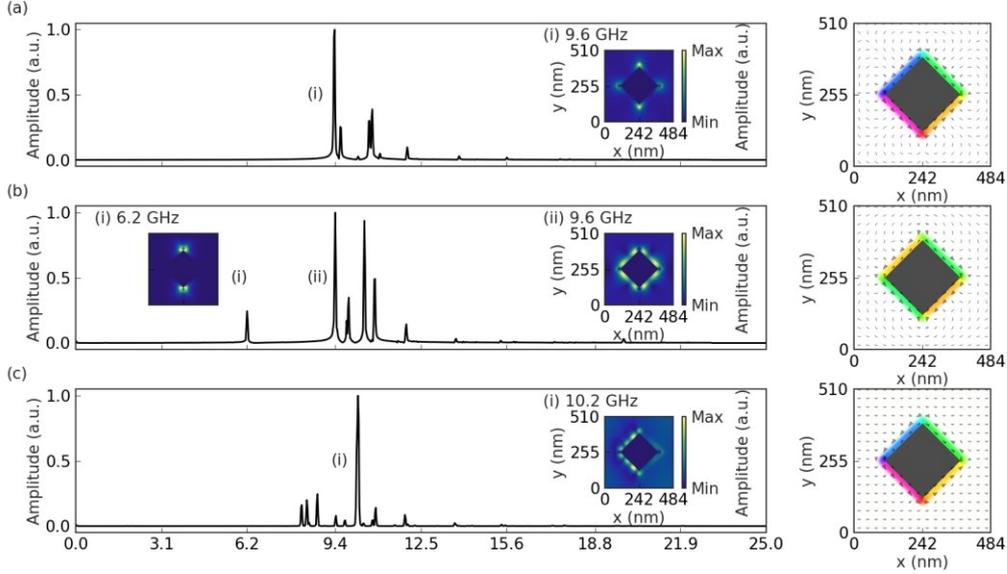

**Fig. 6.** The spectra for the Di-ADL structure with the shell of the $d$ = 20 nm width. (a) The magnetization in the shell forms the closure domain structure, and (b) the head-to-head and tail-to-tail domain walls are present in the bottom and top of the shell. (c) The static magnetic field has been tilted by $10°$ from the normal direction towards the $x$ axis, the magnetization in the shell form the closure domain structure. In simulations the magnetic field of 2.23 Oe was assumed.

**Conclusions**

We have investigated experimentally the SW excitations in ADLs based on Co/Pd multilayers having PMA with antidots of different shapes using TR-MOKE microscope. With the aid of micromagnetic simulations we propose that the emergence of narrow (~10 nm wide) shells with reduced PMA around the antidots are responsible for the observed lower frequency peaks below the fundamental mode frequency, in the SW spectra. The shells with reduced PMA are probably formed due to $Ga^+$ ion irradiation during the patterning process. In those shells the magnetization stabilizes in the film plane, contributing strongly to the measured TR-MOKE signal, which is proportional to the out of plane component of the dynamical magnetization. The different magnetization states in the shells form suitable conditions for formation of different kinds of edge localized SW modes.

We speculate, that the magnetization in the shells around the antidots of the triangle, square and diamond shapes can stabilize in different configurations at remanence, similar to the isolated rings with respective shapes, but having larger width and made of the soft ferromagnetic materials. The variety of the magnetization states in the shells gives opportunity for further exploitation of the edge localized SW spectra, which can be controlled by the magnetization configuration. This can be achieved by developing protocols of the remagnetization with the in-plane component of the external magnetic field, like in the case of the isolated rings based on soft ferromagnetic materials of circular, triangle, square and diamond shapes. [32,33,34] Furthermore, the role of interactions between the edge localized SW excitations, and other types of SWs present in ADL, open the prospect for exploiting collective dynamics in these new kinds of ADLs.

**Acknowledgements:** AB acknowledges S. N. Bose National Centre for Basic Sciences for funding (grant no.: SNB/AB/18-19/211). SM acknowledges DST for INSPIRE fellowship (award number: IF140998). MZ acknowledges support from National Science Center of Poland grant no UMO-2017/27/N/ST3/00419. The simulations were partially performed at the Poznan Supercomputing and Networking Center (Grant No. 398).